\documentclass[twocolumn,floatfix,prl,showpacs,superscriptaddress]{revtex4-1}
\usepackage{graphicx}
\usepackage{amsmath}
\usepackage{color,soul}

\begin{document}
    \title{Nature of Magnetic  Excitations in the High-Field Phase of  $\alpha$-RuCl$_3$}

\author{A.~N.~Ponomaryov}
\thanks{Present Address: Institute of Radiation Physics,  Helmholtz-Zentrum
Dresden-Rossendorf, 01328 Dresden, Germany.}
\affiliation{Dresden High Magnetic Field Laboratory (HLD-EMFL) and
W\"urzburg-Dresden Cluster of Excellence ct.qmat, Helmholtz-Zentrum
Dresden-Rossendorf, 01328 Dresden, Germany}
\author{L.~Zviagina}
\affiliation{Dresden High Magnetic Field Laboratory (HLD-EMFL) and
W\"urzburg-Dresden Cluster of Excellence ct.qmat, Helmholtz-Zentrum
Dresden-Rossendorf, 01328 Dresden, Germany}
\author{J.~Wosnitza}
\affiliation{Dresden High Magnetic Field Laboratory (HLD-EMFL) and
W\"urzburg-Dresden Cluster of Excellence ct.qmat, Helmholtz-Zentrum
Dresden-Rossendorf, 01328 Dresden, Germany}
\affiliation{Institut f\"{u}r Festk\"{o}rper- und Materialphysik,
TU Dresden, 01062 Dresden, Germany}
\author{P.~Lampen-Kelley}
\affiliation{Materials Science and Technology Division, Oak Ridge
National Laboratory, Oak Ridge, TN 37821, USA}
\affiliation{Department of Materials Science and Engineering, University
of Tennessee, Knoxville, TN 37821, USA}
\author{A.~Banerjee}
\thanks{Present Address: Department of Physics and Astronomy, Purdue University, West Lafayette, IN 47907, USA.}
\affiliation{Neutron Scattering Division, Oak Ridge National
Laboratory, Oak Ridge, TN 37831, USA}
\author{J.-Q.~Yan}
\affiliation{Materials Science and Technology Division, Oak Ridge
National Laboratory, Oak Ridge, TN 37821, USA}
\author{C.~A.~Bridges}
\affiliation{Chemical Science Division, Oak Ridge National Laboratory,
Oak Ridge, TN 37821, USA}
\author{D.~G. Mandrus}
\affiliation{Department of Materials Science and Engineering, University of Tennessee, Knoxville, TN 37821, USA}
\affiliation{Materials Science and Technology Division, Oak Ridge
National Laboratory, Oak Ridge, TN 37821, USA}
\author{S.~E.~Nagler}
\affiliation{Neutron Scattering Division, Oak Ridge National
Laboratory, Oak Ridge, TN 37831, USA}
\author{S.~A.~Zvyagin}
\thanks{Corresponding author: s.zvyagin@hzdr.de}
\affiliation{Dresden High Magnetic Field Laboratory (HLD-EMFL) and
W\"urzburg-Dresden Cluster of Excellence ct.qmat, Helmholtz-Zentrum
Dresden-Rossendorf, 01328 Dresden, Germany}

\date{\today}

\begin{abstract}

 We present comprehensive electron spin resonance (ESR)   studies of  in-plane oriented single crystals of $\alpha$-RuCl$_3$, a quasi-two-dimensional material with honeycomb structure,   focusing on its high-field spin dynamics.  The measurements were performed in magnetic fields up to 16 T, applied along  the [110] and  [100] directions. Several ESR modes were detected. Combining our findings with  recent  inelastic neutron- and Raman-scattering data,  
we identify most of the observed excitations.   Most importantly, we  show that   the low-temperature ESR response   beyond the  boundary of the magnetically ordered region  is dominated by  single- and two-particle processes with  magnons as elementary excitations.  The peculiarities of the excitation spectrum in the vicinity of the critical field are discussed.

\end{abstract}

\maketitle

\begin{figure} [!h]

\vspace{0cm}
\hspace*{0cm}
\includegraphics[width=0.47\textwidth]{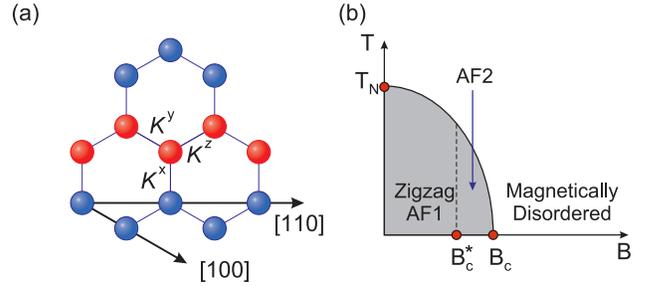}

\caption{\label{fig:Structure} (a) Schematic view of the honeycomb
structure,  showing the [100] and [110] axes relative to the Ru--Ru bonds. Ru
ions from adjacent zigzag chains are shown by different colors.
(b) Schematic temperature-field phase diagram for $\alpha$-RuCl$_3$. 
 AF1 and AF2 correspond to different  low-temperature antiferromagnetically  ordered  phases.  
 }
\end{figure}

Spin systems with honeycomb structures have recently attracted a great
deal of attention, in particular in connection with the 
Kitaev-Heisenberg
model \cite{Chaloupka}. The model predicts a variety of magnetic
phases, ranging from the conventional N\'{e}el state to a  quantum spin
liquid, with the excitation spectrum  formed by spin-flip excitations, fractionalized into gapped flux excitations 
and gapless Majorana fermions \cite{Kit_1}.  $\alpha$-RuCl$_3$
has been proposed as one of the prime candidates to test this model
\cite{Plumb}.    In this material,   the multiorbital
$5d$ $t_{2g}$ state can be mapped into a single orbital state with 
effective pseudospins $j_{eff}$ = 1/2. The spins are arranged into
a two-dimensional (2D) honeycomb lattice [Fig.~\ref{fig:Structure}(a)]
with bond-dependent interactions,  defined by the
Kitaev parameter $K$ in the Hamiltonian:

\begin{equation}
\begin{aligned}
\label{Ham}  \mathcal{H}=\sum_{ \langle ij \rangle}
\bigg[J\pmb{S}_i\cdot \pmb{S}_j + KS_i^\gamma S_j^\gamma +
\Gamma\Big(S_i^\alpha S_j^\beta + S_i^\beta S_j^\alpha\Big)\bigg]-
\\
- \mu_B \sum_i \textbf{B}\cdot \textbf{g} \cdot \pmb{S}_i.
\end{aligned}
\end{equation}
Here,   $S_i$ and  $S_j$  are spin-1/2 operators at site $i$ and $j$,
respectively, $J$ is the Heisenberg exchange parameter, $\Gamma$ represents a
symmetric off-diagonal term, and $\mu_B$,  $\textbf{B}$, $\textbf{g}$ 
correspond to the Bohr magneton,  magnetic field, and
  $g$ tensor,  respectively ($\alpha$ and $\beta$ are perpendicular to the Kitaev spin axis
$\gamma$). A number of  sets of parameters of the generalized Kitaev-Heisenberg
model for  $\alpha$-RuCl$_3$ have been proposed (for a review see, e.g., Ref. \cite{Chern}). Below $T_N\sim 7$ K,  the system undergoes the transition into 3D long-range zigzag magnetically ordered phase \cite{Sears} 
associated with a triple-layer structure modulation in the direction perpendicular to the honeycomb direction [phase AF1 in  Fig.~\ref{fig:Structure}(b)]. Magnetic ordering is suppressed under the  application of magnetic field  $B_c$ \cite{Johnson}, followed by 
a partial polarization of the ground state \cite{Rem_FSPS}. In addition, a signature of the second phase (AF2) has been detected   between $B_{c}^*$ and $B_{c}$ \cite{Ban_2,LK_1,Balz}.   For   $H \parallel [110]$,  the critical fields are $B_{c}^*=6.1$ and $B_{c}=7.3$ T,  and these converge at $B_{c}= 7.6$ T as one moves toward $H \parallel [100]$, where the separation is small, if not zero  \cite{LK_1}.


One striking peculiarity of the spin dynamics in $\alpha$-RuCl$_3$ is
the presence of a broad excitation continuum, 
which has been interpreted   as a potential signature
of fractionalized Majorana excitations \cite{San,Ban,Do,Ban_3}. It  
can be observed  up to well above 100 K,
indicating the rather high-energy scale of  magnetic interactions
in this compound.   The continuum  remains present well  below  $T_N$, when the ground state is
magnetically ordered  and the
low-energy excitation spectrum is formed by two antiferromagnetic resonance (AFMR)
modes \cite{Zvyagin_ARC,Little,Wu}.
Based on that, $\alpha$-RuCl$_3$ was proposed  to be in close proximity to the predicted Kitaev quantum spin liquid \cite{Ban}.  In the field-induced  disordered phase the continuum is also present and
gapped, with the gap gradually increasing with the applied
magnetic field \cite{Zhe,Wellm,Ban_2,Loosd,Lemm}. 

Alternatively, the continuum
can be  described  in terms of  incoherent multimagnon
processes \cite{Winter,Winter_a}. In line with that, recent calculations \cite{Chern} 
 point toward the physics of the strongly interacting and mutually decaying magnons, not to that of the fractionalized 
excitations.

Recent high magnetic field spectroscopy measurements revealed a very rich excitation spectrum in the field-induced magnetically disordered phase \cite{Zvyagin_ARC,Zhe,Wellm,Loosd,Lemm}, including several modes below the continuum.  The remarkably large slope of some of them implies  the presence of transitions with $\Delta S=2$ (contrary to $\Delta S=1$, expected for conventional one-particle excitations in  $S=1/2$ systems) \cite{Zvyagin_ARC}. This observation  strongly suggested  that the high-field spin dynamics in $\alpha$-RuCl$_3$   has an emergent multiparticle nature,  raising an important  question on the nature of the observed excitations.

 To
understand the complex spin dynamics in $\alpha$-RuCl$_3$, a
comparative analysis of available experimental data is essential. 
Unfortunately, one critical
shortcoming of the majority of  magnetic   studies of $\alpha$-RuCl$_3$
comes from ignoring its in-plane anisotropy, which makes such a comparison 
challenging or even impossible. 
The  anisotropy appears to be rather pronounced, as followed from  
high-field electron spin resonance
  \cite{Zvyagin_ARC} and magnetic susceptibility \cite{LK_1,LK_2}
measurements, suggesting the presence of the Kitaev parameter $K$ and
     symmetric off-diagonal spin exchange $\Gamma$ [Eq. (\ref{Ham})] as
two key  sources of the anisotropy  \cite{LK_2,Suzuki}. 

Electron spin resonance (ESR) is traditionally  recognized as one of the most sensitive  high-resolution spectroscopy  tools for studying the spin dynamics  in strongly correlated electron systems, capable of probing  not only conventional magnons,  but also fractional excitations 
(such as spinons and  solitons \cite{Asano,Sol,Povarov,CPC}),  the property of  magnetic materials   with  quantum spin liquid ground states. Here, we present results of high-field   tunable-frequency ESR studies  of 
$\alpha$-RuCl$_3$, focusing on its spin dynamics in the field-induced magnetically disordered phase. 

The  measurements were performed on high-quality single crystals from the same batch as reported previously  \cite{Zvyagin_ARC}. 
The platelike samples were
prepared using a vapor-transport technique starting from pure RuCl$_3$
powder   and have typical sizes of 3x3x0.5 mm$^3$.  
The experiments 
were performed  employing a 16 T transmission-type multifrequency
ESR spectrometer,  similar to that described in Ref.
\cite{Zvyagin_INSR}. A set of backward-wave oscillators, Gunn diodes,
and VDI
 microwave sources (Virginia Diodes Inc, USA) was used,
allowing us to study  magnetic excitations 
in a  broad quasicontinuously covered frequency range,  from 0.05
to 1.2 THz (corresponding to an energy range of about 0.2-5 meV). The experiments were performed 
in the Voigt configuration 
with magnetic fields $H \parallel [100]$ and $H \parallel [110]$
[i.e., applied parallel and  perpendicular to a Ru--Ru bond direction,
respectively, Fig.~\ref{fig:Structure}(a)]. Fine orientation of the samples was done  $in~situ$, employing a goniometer with the rotation axis normal to honeycomb layers.  
A wire-grid polarizer was installed just before the sample, allowing
us to select the polarization of the incident THz radiation with  respect
to the applied magnetic field and crystallographic axes. 

\begin{figure} [t]

\begin{center}
\vspace{0mm}
\hspace*{-0.4cm}
\includegraphics[width=0.54\textwidth]{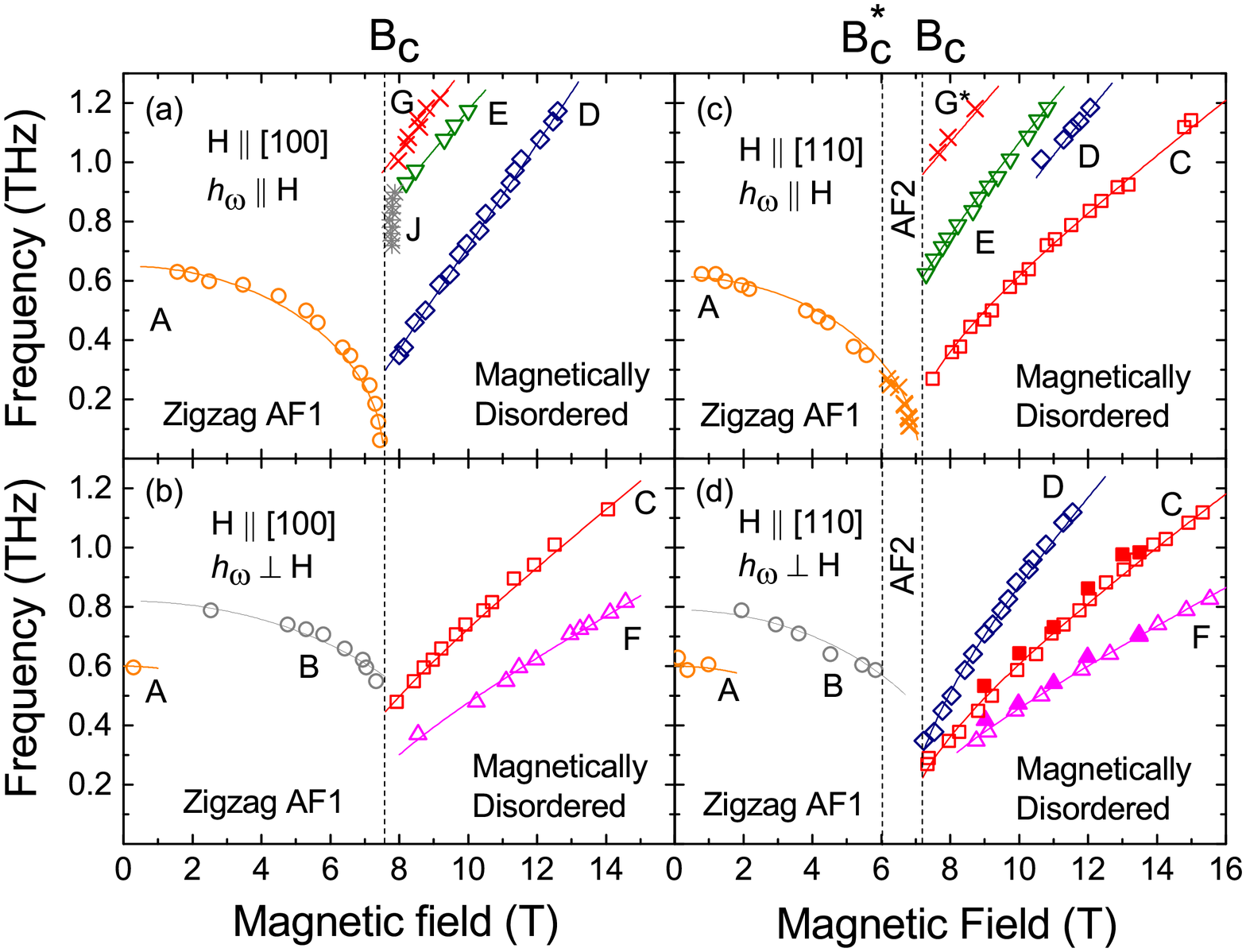}
\vspace{-4mm}
\caption{\label{fig:FFD}   Frequency-field dependences of magnetic
excitations in $\alpha$-RuCl$_3$   for $H \parallel [100]$ (a, b) and  $H
\parallel [110]$  (c, d) ($h_{\omega} \parallel H$ (a, c),   $h_{\omega} \perp H$ (b, d), where $h_{\omega}$ is the magnetic component of the THz  radiation; 
$T=1.5$ K).  Absorptions denoted by crosses  were observed using unpolarized  
radiation. Solid lines are guides to the eye. The vertical dashed lines indicate critical fields as determined in Ref. \cite{LK_1}. 
The closed squares and triangles in (d) denote inelastic neutron-scattering data from Ref. \cite{Balz}. }
\end{center}
\end{figure}

The 
frequency-field dependences of polarized ESR in
$\alpha$-RuCl$_3$ for $H \parallel [100]$  and  $H \parallel [110]$  
 are shown in 
Fig.~\ref{fig:FFD}. Some examples of polarized-ESR measurement data are shown in the Supplemental Material \cite{SM}. 
 Two modes  were observed in the low-field zigzag ordered phase, as reported  previously  \cite{Zvyagin_ARC}.
These excitations (modes $A$ and $B$)  correspond to conventional relativistic AFMR modes 
excited at the $\Gamma$ point,  with the zero-field frequencies 0.62
and 0.8 THz \cite{Wu}. Both modes exhibit pronounced softening in 
magnetic field.  The low-frequency AFMR 
mode $A$ is dominantly excited when  $h_{\omega} \parallel H$, while the mode $B$
corresponds to excitations
with $h_{\omega} \perp H$. At low fields,  mode  $A$ was observed also when $h_{\omega} \perp H$. 
Such an unusual behavior can be explained by a change in domain populations,  as suggested  by  neutron-scattering studies \cite{Ban_2}. 
 Detailed spin-wave-theory analysis of the AFMR
spectrum (including the polarization dependence of magnetic excitations) was performed by Wu $et~al.$ \cite{Wu}, revealing an overall good  agreement with the obtained   experimental data.

\begin{figure} [!h]

\begin{center}
\vspace{2mm}
\hspace*{-0mm}
\includegraphics[width=0.49\textwidth]{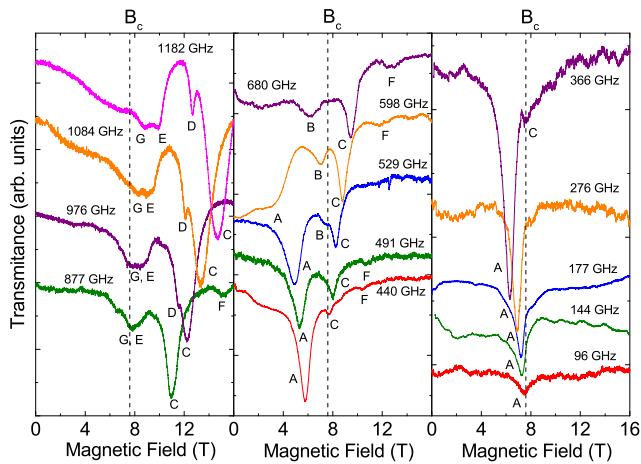}
\vspace{0mm}
\caption{\label{fig:SP_UNP_100}   Examples of unpolarized ESR spectra in $\alpha$-RuCl$_3$   for $H \parallel [100]$ at various frequencies; $T=1.5$ K. 
The spectra are normalized by the zero-field transmittance background  and offset for clarity. 
The vertical line indicates the critical field as determined in Ref. \cite{LK_1}.}
\end{center}
\end{figure}

\begin{figure} [!h]
\begin{center}
\vspace{2mm}
\hspace*{-0mm}
\includegraphics[width=0.49\textwidth]{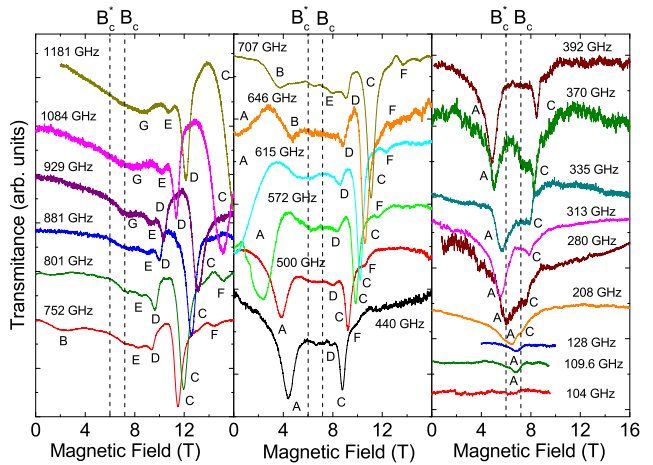}
\vspace{-0mm}
\caption{\label{fig:SP_UNP_110}   Examples of unpolarized ESR spectra in $\alpha$-RuCl$_3$   for $H \parallel [110]$ and $T=1.5$ K.  The spectra are normalized by the zero-field transmittance background  and offset for clarity.  The vertical lines indicate critical 
fields as determined in Ref. \cite{LK_1}.}
\end{center}
\end{figure}

For both orientations of  applied magnetic field,  the intensity of the AFMR  modes decreases significantly when approaching the critical region.  These changes can be particularly well seen in unpolarized ESR spectra  [Figs.~\ref{fig:SP_UNP_100} and \ref{fig:SP_UNP_110}]. Remarkably, our ESR measurements revealed the presence of  AFMR mode $A$  not only below $B_{c}^*$, but also  between  $B_{c}^*$ and $B_{c}$ [Fig.~\ref{fig:FFD}(c)].  It has been recently proposed,  that the antiferromagnetic interlayer coupling in $\alpha$-RuCl$_3$
results in  a triple-layer structure modulation in the direction perpendicular to the honeycomb direction (corresponding to the magnetic order 3f-zz), while the ferromagnetic interaction  would lead to zigzag ordered  state with a unit cell of six layers (6f-zz); the latter  is likely realized in $\alpha$-RuCl$_3$  between  $B_{c}^*$ and $B_{c}$  \cite{Janssen}.    Based on this assumption, the  observation of the mode $A$ in the AF2 phase suggests  the coexistence of the 3f-zz and 6f-zz magnetic structures in this narrow intermediate field range   [Fig.~\ref{fig:Structure}(b)]. More details of high-field magnetic structure studies of  $\alpha$-RuCl$_3$  will be reported elsewhere \cite{Balz_2}.

Several magnetic resonance modes  were observed above $B_{c}$.  The frequency-field diagrams of these  modes for different polarizations of the incident  THz radiation  are shown in Fig.~\ref{fig:FFD}.  

\begin{figure} [!h]
\begin{center}
\vspace{0mm}
\hspace*{-0.7cm}
\includegraphics[width=0.49\textwidth]{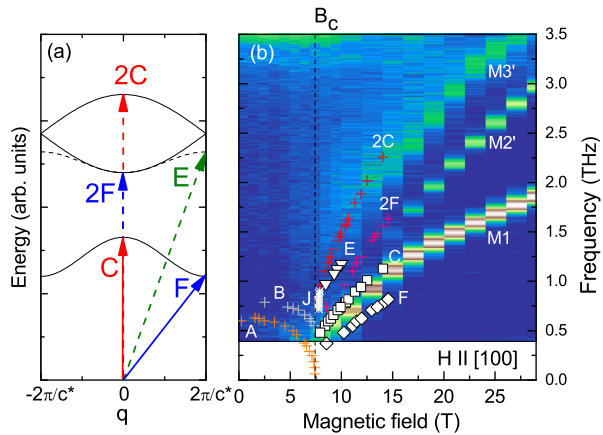}
\vspace{-2mm}
\caption{\label{fig:Disp} (a) Proposed schematic energy diagram for $\alpha$-RuCl$_3$ in an arbitrary magnetic field above $B_{c}$. 
The modes $C$ and $F$ are single-magnon excitations, while the modes $2C$ and $2F$ correspond to two-magnon excitations.  The mode $E$ corresponds to an  excitation of a two-magnon bound state. (b) 
Frequency-field dependences of selected  ESR modes  [from  Figs.~\ref{fig:FFD}(a) \ref{fig:FFD}(b)] and the color contour plot of the high-field Raman-scattering intensity \cite{Lemm} ($H \parallel [100]$).} 
\end{center}
\end{figure}

Recent  neutron-scattering measurements of $\alpha$-RuCl$_3$  revealed a sharp magnon mode at the lower bound of a strong continuum \cite{Balz}. This mode has a measurable  dispersion in the direction perpendicular to the honeycomb planes,  suggesting the presence of non-negligible interplane interactions (the dispersion perpendicular to the plane  was seen also in the magnetically ordered phase below $B_{c}$, but it is much weaker than the in-plane dispersion).   The corresponding neutron-scattering data at (0, 0, 3.3)  and (0, 0, 4.3)  are shown in Fig.~\ref{fig:FFD}(d) by closed squares and triangles, respectively.  The dispersion  periodicity  along the (0, 0, $L$) direction  suggests that the excitation energy at the $\Gamma$ point (maximum of the excitation dispersion) and at the  magnon zone boundary (dispersion minimum) are approximately the same as  for (0, 0, 3.3)  and (0, 0, 4.3), respectively.  Based on that, the excitations  $C$ and $F$ are identified as relativistic and exchange modes of magnetic resonance [Fig.~\ref{fig:Disp}(a)]  (the same interpretation is  given  in Ref. \cite{Chern}); similar excitations were  observed, e.g., in the field-induced polarized phase in the  triangular-lattice antiferromagnet Cs$_2$CuCl$_4$  \cite{Zvyagin_CCB}. The mode $C$ is the most intensive resonance 
(Figs.~\ref{fig:SP_UNP_100} and \ref{fig:SP_UNP_110}),  having maximal intensity for the polarization  $h_{\omega} \perp H$.
This mode was observed also  by means of far-infrared \cite{Zhe, Loosd} and Raman-scattering \cite{Lemm} spectroscopy   [the latter is denoted as $M1$ in Fig.~\ref{fig:Disp}(b)]. 
  The mode $F$ has a polarization  $h_{\omega}\perp H$.  The corresponding transitions (modes $C$ and $F$) are shown in Fig.~\ref{fig:Disp}(a)   by the solid red and blue arrows, respectively.  
The observation  of the exchange mode $F$ (which is, as expected,  much weaker than the mode $C$)  becomes  possible due to the  staggered Dzyaloshinskii-Moriya interaction \cite{Bar, Sakai}, which is allowed in $\alpha$-RuCl$_3$ due to the absence of an inversion symmetry center between the Ru ions in adjacent  layers. Our scenario is supported by recent calculations for a three-dimensional 
exchange model \cite{Janssen}.  The distance between the $C$ and $F$ modes gets larger with increasing field, indicating that spin correlations in the system are becoming  less 2D in high fields.  Similar behavior was  observed  by inelastic neutron-scattering experiments \cite{Balz}.

Raman scattering is known as a very powerful tool to probe two-particle processes in strongly correlated spin systems. 
Such two-magnon excitations, the modes $M3'$ and $M2'$,  were observed in $\alpha$-RuCl$_3$ in the field-induced phase \cite{Lemm,rem2}
(Fig.~\ref{fig:Disp}(b); for comparison we show simulated modes $2C$ and $2F$ with the excitation energy twice larger than that for the modes C and F, respectively). The  continuum is spread well above $2C$, suggesting contributions of multiple-particle   processes in  the entire Brillouin zone \cite{Janssen}.  Based on the proposed scenario, one would expect the presence of higher-energy   excitations (such as modes $G$ and $G*$ in  Fig.~\ref{fig:FFD}), involving  multiparticle  processes with different wave numbers.

The ESR mode E,  with   excitation energy slightly larger  than that for the mode 2F (but smaller than that for the mode 2C), was observed for $h_{\omega} \parallel H$ and can be tentatively interpreted  as an excitation of a two-magnon bound state [dashed green arrow in   Fig.~\ref{fig:Disp}(a)].  

In the vicinity of $B_c$ the modes $G$ and $E$ are superimposed (Fig.~\ref{fig:SP_UNP_100}). To  obtain more details about  this critical range, we refer to our polarized ESR measurement data (Supplemental Material \cite{SM}). Surprisingly, at about $B_c$  our experiments revealed a broad dip denoted as $J$, whose field position is almost independent on the frequency.  The dip was observed in the $\sim700-900$ GHz frequency range with the polarization of the incident THz radiation  $h_{\omega} \parallel H$ ($H \parallel [100]$, Fig. 1(a), Supplemental Material) \cite{rem3}. The position of the dip $J$ is shown in Figs.~\ref{fig:FFD}(a) and ~\ref{fig:Disp}(b). Remarkably, this frequency range is located between the excitation energies for modes $2F$ and $2C$, corresponding to the lower and upper boundaries of the two-magnon continuum, respectively (Fig.~\ref{fig:Disp}).
This strongly suggests that the field-induced transition from the magnetically ordered to disordered phase strongly affects not only the ground state properties, but also the excitations spectrum, including  multiparticles processes. We hope that our observation will stimulate further theoretical studies of the unconventional spin dynamics   in $\alpha$-RuCl$_3$, in particular, in the critical regime in the vicinity of $B_c$.

The ESR mode $D$ is relatively weak at low frequencies,
gaining  intensity at higher  frequencies  and fields. This mode is excited for both
polarizations of  incident THz radiation,  $h_{\omega} \parallel H$  and
$h_{\omega} \perp H$   (Fig.~\ref{fig:FFD}). Similar to other high-field modes,   the resonance field for the mode D exhibits a 
$60^{\circ}$ periodicity \cite{Zvyagin_ARC}.
 On the other hand,    the angular dependence of this mode   is
significantly different from the others (e.g., for modes $C$, $E$, $F$), demonstrating a shift of 
$30^{\circ}$.   The observed very peculiar angular dependence of  mode $D$  might  provide  a potential hint for identifying the nature  of this excitation.

Very recently, a plateau in the  thermal Hall effect has been observed over a finite field range \cite{Kasahara,Yokoi,Yam}.   This has been interpreted as a signature of  fractional  non-Abelian excitations, possibly the Majorana fermions of the Kitaev model on a honeycomb lattice.  The presence of a plateau over a limited range of applied fields (approximately between 9.7 and 11.5 T for  $H\parallel [110]$ \cite{Yokoi}) would suggest the presence  of additional phase transitions at the fields corresponding to the upper and lower bounds of the plateau.  Possible evidence for that has been seen in magnetocaloric \cite{Balz} and magnetostriction \cite{Gass} experiments, while another thermodynamic study (magnetic Gr\"{u}neisen parameter  and specific  heat) detected no sign of such transitions  \cite{Bach}. Our  high-field ESR measurements show  magnon  modes, characteristic of a partially polarized state  emerging $right~above$ $B_c$,   not revealing  any evidence  for additional high-field phases or phase transitions in magnetic fields  up to 16 T.  The question of such a coexistence (the nontrivial topological excitations, if any, and conventional bulk magnons,  observed by us  in $\alpha$-RuCl$_3$)  remains open, demanding more systematic experimental and theoretical investigations.

In conclusion, we have reported on the  high-resolution  high-field  THz ESR spectroscopy  studies of in-plane oriented  single crystals of  $\alpha$-RuCl$_3$ in magnetic fields up to $B_c$ and beyond, applied  parallel and  perpendicular to  Ru--Ru bond directions. 
 We have confirmed the rather anisotropic ESR response, 
highlighting the significant  role  of  anisotropic in-plane interactions in $\alpha$-RuCl$_3$. Complemented  by the  results of recent inelastic neutron- and Raman-scattering measurements, we have argued that the high-field spin dynamics in this material is dominated by one- and two-particle excitations identified as magnons. We hope that our observations will stimulate further theoretical studies of the unconventional spin dynamics   in $\alpha$-RuCl$_3$, in particular, in the critical regime in the vicinity of $B_c$.

This work was supported by the Deutsche Forschungsgemeinschaft  through Garnt No.ZV 6/2-2, the excellence
cluster \textit{ct.qmat} (EXC2147, Project-id No 390858490), and SFB 1143, as well as by the HLD at HZDR, member of the European Magnetic Field Laboratory. A.B.  and S.E.N.  were supported
by the Division of Scientific User Facilities, Basic Energy Sciences
US DOE, P.L.-K. and D.G.M. by the Gordon and Betty
Moore Foundation's EPiQS Initiative through Grant GBMF4416, J.-Q.Y.
and C.A.B. by the U.S. Department of Energy, Office of Science,
Office of Basic Energy Sciences, Materials Sciences and Engineering
Division. We would like to thank D. Wulferdung and P. Lemmens  for sharing their experimental data. We acknowledge discussions with M.~Vojta, A. ~Chernyshev, M.~Zhitomirsky, and A.~ Kolezhuk.

\begin{center}
\Large{\textbf{Supplemental Material}}
\end{center}

\begin{figure} [h!]

\begin{center}
\vspace{-0.5cm}
\hspace*{-0.9cm}
\includegraphics[width=0.55\textwidth]{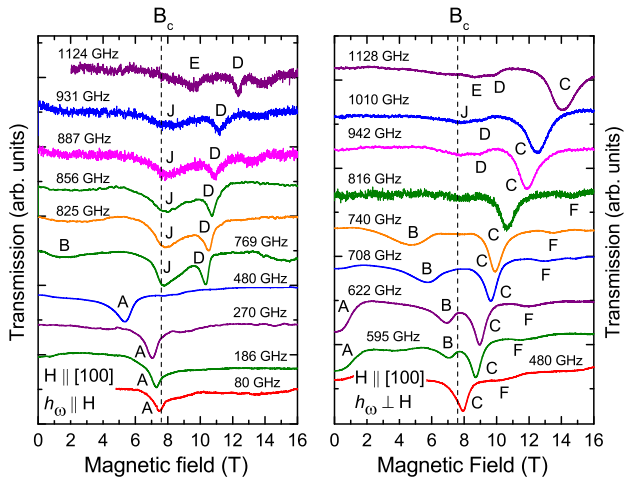}
\vspace{0cm}
\caption{\label{fig:SM_1}  Examples of polarized ESR spectra in $\alpha$-RuCl$_3$   for $h_{\omega} \parallel H$ (a) and   $h_{\omega} \perp H$ (b) ($h_{\omega}$ is the magnetic component of the THz  radiation).  $H \parallel [100]$,  $T=1.5$ K.  The spectra are normalized by the zero-field transmittance background  and offset for clarity. 
The vertical line indicates the critical field as determined in Ref.~\cite{LK_1}.
 }
\end{center}
\end{figure}

\begin{figure} [h!]

\begin{center}
\vspace{-0.5cm}
\hspace*{-0.9cm}
\includegraphics[width=0.55\textwidth]{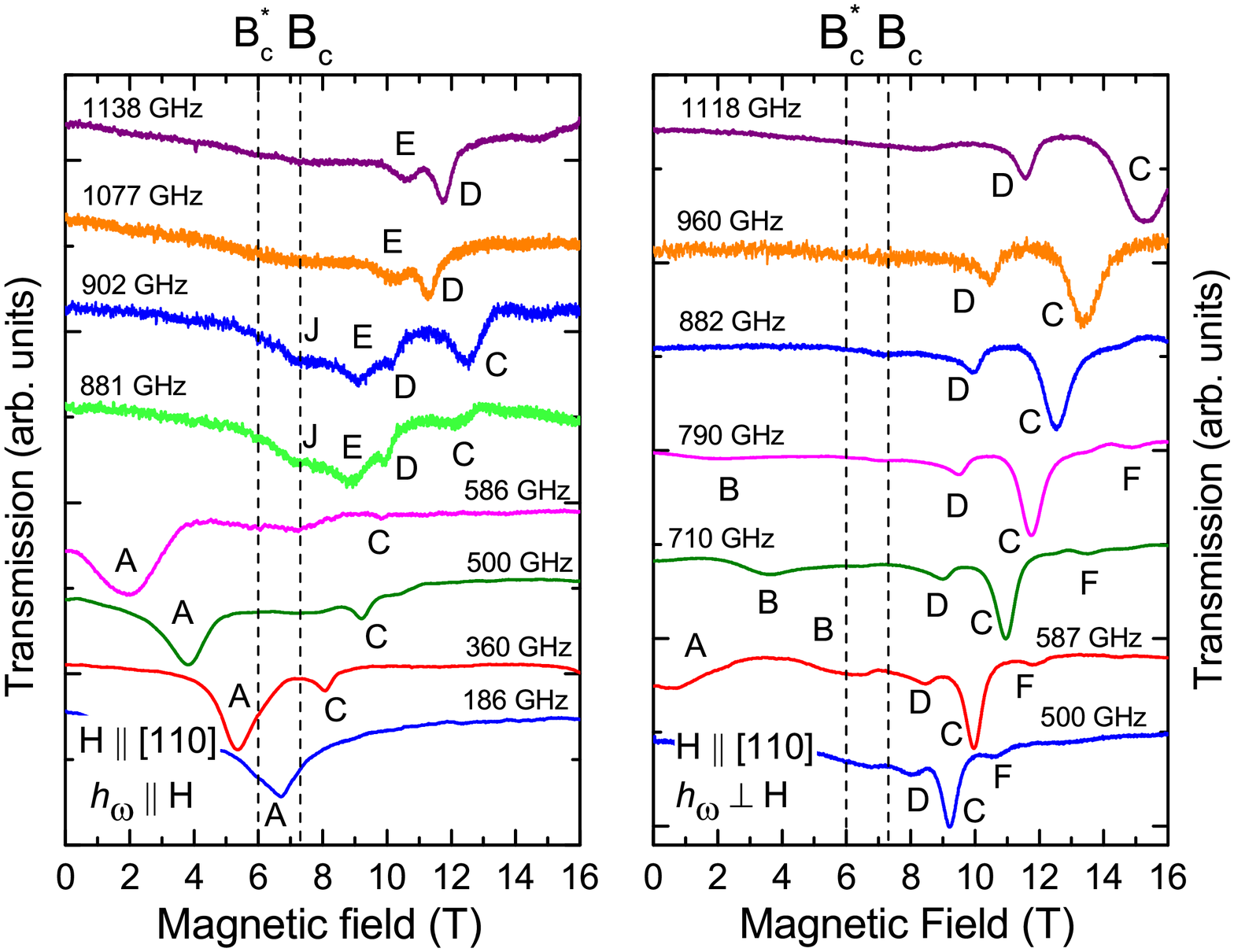}
\vspace{0cm}
\caption{\label{fig:SM_2}  Examples of polarized ESR spectra in $\alpha$-RuCl$_3$   for $h_{\omega} \parallel H$ (a) and   $h_{\omega} \perp H$ (b) ($h_{\omega}$ is the magnetic component of the THz  radiation).  $H \parallel [110]$,  $T=1.5$ K.  The spectra are normalized by the zero-field transmittance background  and offset for clarity. 
The vertical lines indicate the critical fields as determined in Ref.~\cite{LK_1}.
 }
\end{center}
\end{figure}


\begin{thebibliography}{99}
\bibitem{Rem_FSPS} Because of the presence of the finite Kitaev coupling,  a full spin polarization can be achieved  only at infinitely large magnetic field.
\bibitem{rem2} Contrary to our interpretation, the mode  $M2'$  in Ref. \cite{Lemm} is  identified   as excitation of a two-magnon bound state.
\end{thebibliography}
\end{document}